\documentclass[nofootinbib,preprint,superscriptaddress]{revtex4}
\usepackage{amsmath}
\usepackage{amssymb}
\usepackage{graphicx}

\usepackage{color}

\begin{document}

\title{Asymmetric space-dependent systems:
  Partial stabilization through the addition of noise and exact
  solutions for the corresponding nonlinear Langevin equations }

 \author{Kwok Sau Fa
  \footnote{Permanent address:
    Department of Physics, Universidade Estadual de Maring\'{a},
    Av. Colombo 5790, 87020-900, Maring\'{a}-PR, Brazil}
\footnote{corresponding author email: kwok@dfi.uem.br}}
\affiliation{Departmento de Física, Universidade Federal do
  Paran\'{a}, 81531-980 Curitiba-PR, Brazil}

\author{Choon-Lin Ho}
\affiliation{Department of Physics, Tamkang University,
  Tamsui 25137, Taiwan}

\author{Y. B. Matos}
\affiliation{Departmento de Física,
  Universidade Federal do Paran\'{a}, 81531-980 Curitiba-PR,
  Brazil}

\author{M. G. E. da Luz
  \footnote{email: luz@fisica.ufpr.br}}
\affiliation{Departmento de Física,
  Universidade Federal do Paran\'{a}, 81531-980 Curitiba-PR,
  Brazil}



\begin{abstract}
  In many instances, the dynamical richness and complexity
  observed in natural phenomena can be related to stochastic
  drives influencing their temporal evolution.
  For example, random noise allied to spatial asymmetries may
  induce stabilization of otherwise diverging trajectories in
  dynamical systems.
  However, to identify how exactly this takes place in actual
  processes usually is not a simple task.
  Here we unveil a few trends leading to dynamical stabilization
  and diversity of behavior by introducing Gaussian white
  noise to a class of exactly solvable non-linear deterministic
  models displaying space-dependent drifts.
  For the resulting nonlinear Langevin equations, the
  associated Fokker-Planck equations can be solved through
  the similarity method or the Fourier transform technique.
  By comparing the cases with and without noise, we discuss
  the changes in the systems dynamical characteristics.
  Simple examples of drift and diffusion coefficients are
  explicitly analyzed and comparisons with some other
  models in the literature are made.
  Our study illustrates the rich phenomenology originated from
  spatially heterogeneous dynamical systems under the influence
  of white noise.
\end{abstract}

\maketitle


\section{Introduction}

The great assortment of responses to external stimuli is one
of the key factors generating the behavioral diversity common
to many natural phenomena \cite{nakamura-1997,page-2011}.
For instance, this can give rise to the emergence of complexity
and spatial-temporal patterns in a broad range of processes
\cite{rand-1994,rand-1995}, even if they are restricted to certain
constraints (say, having to follow a gradient flow) as in the
evolution of coarsening systems \cite{cugliandolo-2015,mayer-2004}.

There are distinct features allowing for such variety of
evolution trends (for a review see, e.g., \cite{stankovski-2017}).
But certainly, spatial heterogeneities and/or asymmetries are among
the most ubiquitous ones 
\cite{agafonov-2005,cumming-2011,lowery-2018,edri-2020}.
Actually, effects like spatial-temporal oscillations \cite{krause-2018},
resonances \cite{edri-2020-physicad} and strong dispersion
\cite{sun-2021} can all be triggered by inhomogeneous environments.
This is particularly true in biologically-related problems
where space-dependent coefficients may greatly influence the
variability of population genetics \cite{kimura,magin,kwok4}
and growth \cite{calisto2,popul,kwok3,richa,saka,goel,roman,gomp},
the type of diffusion across membranes \cite{sen}, and the onset
of anomalous mobility \cite{wu}, just to cite a few examples.
It is also needless to emphasize that landscape profile changes
are fundamental to understand large and strongly correlated
systems, as in ecology \cite{rand-1995,cumming-2011,lowery-2018}.

Nonetheless, as relevant as to induce distinct comportment,
spatially asymmetric interactions and drives have a crucial role
in stabilizing, synchronizing and promoting cooperative feedback
\cite{rubchinsky-1999,bragard-2001,russo-2011,nicolaou-2021,
wang-2021}.
On the one hand, these characteristics are essential to maintain
the functional diversity in the natural world
\cite{pikovsky-2001,boccaletti-2002,mokekilde-2002,strogatz-2004,
barabas-2022}, preventing trivial dynamics \cite{stankovski-2017}.
On the other hand, the existence of a phase or state space
displaying multi-stability \cite{manchein-2017,silva-2017}
(conceivably created by heterogeneous media) is not enough to
avoid trivial (stable) attractors \cite{dudkowski-2016}.
Thence, escaping or switching mechanisms from dynamical traps
\cite{zaslavsky-2002,manrubia-2004} are commonly found in
systems presenting diversity of behavior, notably in
non-equilibrium as well as in complex systems \cite{yam-1997}.
Given that randomness is rather effective in generating such
types of mechanisms \cite{arnold-2002,freidlin-2012}, the
somewhat omnipresence of stochasticity in a huge number of
physical phenomena is far from being a surprise
\cite{nakamura-1997,page-2011,strogatz-2004,freidlin-2012}.

The above discussion supports the recognized significance of
the generalized Langevin equation (GLE) in describing countless
realistic processes, appealing rather directly to our intuition
(for an overview see \cite{ernesto-2020} as well the refs.
therein).
Broadly, it combines the deterministic Newton's second law with external
stochastic forces
\cite{risken,gardiner,coffey,gitter,snook,kwoklivro,nad,li}. 
However, the GLE is a stochastic differential equation, 
often being difficult to treat for arbitrary drifts and random
forces.
One possible approach is then to convert the GLE into the
Fokker-Planck equation, determining an associated probability
density function (PDF) for relevant quantities \cite{risken,st,ya,fe,ja}.
In this way, one can exploit a large number of methods available
in the literature for solving differential equations.  
This constitutes a traditional framework to tackle innumerous
problems, particularly those whose physical parameters 
depend non-trivially on space.
As illustrations we mention the modeling of: transport processes
\cite{zop,zop2}, organic semiconductors \cite{rais},  star-shaped polymer translocation into a nanochannel \cite{mesay},
periodic porous material \cite{dudko}, and turbulent two-particle
diffusion in configuration space
\cite{richar,richar2,richar3,richar4,richar5}.
Further, in the case of drifts with time-dependent coefficients
(even if implicitly), the Fokker-Planck equation approach helps
to understand unusual dynamics, as the asymptotic of continuous
time random walk models \cite{kwok1,kwokPhysScr2,kwokPhysScr3}, logarithmic oscillations
for moments of physical variables \cite{kwok2} and dynamical
diversity for systems driven by colored noise
\cite{schi,seki,wang,calisto,aquino,tao,kwokload,kwok2019,kwokPhyScr,kwok2020}.

In this contribution we shall address the interplay between
deterministic and stochastic drives in establishing time
evolution traits.
We discuss a way to avoid (1) steadily stopping and (2)
monotonic diverging orbits by adding noise to a class of
spatially asymmetric problems, partially stabilizing the
systems \cite{manchein-2017} and thus allowing richer dynamics
\cite{nakamura-1997,page-2011,stankovski-2017}.
We note there is a vast literature rigorously classifying richness
(i.e., diversity of behavior) in dynamical systems, see e.g.
\cite{bonatti-2006}.
Here we assume a straightforward point of view.
So, by richer we just means to have arbitrary (eventually
involved and irregular \cite{bonatti-2006,afraimovich-2003})
trajectories, but precluding the above asymptotic tendencies
(1) and (2).

To keep the problem as simple as possible, although displaying
spatial heterogeneity, we suppose a set of one-dimensional
first-order differential equations, whose drift coefficients
depend on the sign of their dependent variable $x(t)$ (see
next Section).
Given their functional form $dx(t)/dt = F(x;\Lambda)$
--- with $\Lambda$ representing the collection of parameters ---
straightforward dynamics is simple to identify.
Indeed, they correspond to $dx(t)/dt = 0$ for some finite
time $t=\tau$, case (1), or $dx(t)/dt > 0$ ($dx(t)/dt < 0$)
for all $t$, case (2).
For our prototype models we first show that the pure deterministic
evolution tends to be rather trivial in the aforementioned
sense for a very large region of the $\Lambda$ space.
Then, we include into the equations a generic multiplicative
noise term driven by Gaussian white noise under the
Stratonovich prescription.
This yields nonlinear Langevin equations, displaying anomalous
diffusion.

Following a previously developed method \cite{kwok5} and a
transformation scheme amenable to problems possessing scaling
similarity \cite{kwoklivro,bc,ho1,ho2,ho3,ho4}, we are able to
exactly solve the related Fokker-Planck equation for
a considerably large range of parameter values.
For some instances where such prescription does not work, we
use the Fourier transform technique.
From the analytic solutions we analyze the difference between
the evolution with and without the stochastic component.
In particular, we examine how the stochasticity evades
dynamical steady behavior, partially stabilizing the systems, and also how the emerging evolution
diversity depends on the specific regions of the $\Lambda$
space.

Finally, concrete simple examples are explored in more details.
The resulting PDFs are studied for some parameter values
and a few cases are compared with related traditional models in
the literature.

\section{The set of models}

As stated in the introduction section, our goal is to unveil
potential mechanisms (based on the addition of noise) preventing
the systems to go into trivial time evolutions, as the previously
mentioned instances (1) and (2).
So, we shall work directly with the systems velocity $dx/dt$ and
thus to consider first-order differential equations, relevant in
distinct problems where the interest relies on the configuration
space dynamics \cite{llibre-2023}.
We emphasize that our focus here is not in any specific
physical system.
Therefore, our choice of $F$ in $dx(t)/dt = F(x;\Lambda)$ below
is such that it: does present asymmetric spatial dependence,
in certain conditions can give rise to dynamical richness,
is amenable to analytic solutions and finally, for certain
values of the parameters recovers known models in the literature.

\subsection{Deterministic dynamical system}
\label{sec:model-det}

Consider a one-dimensional dynamical variable $x(t)$, whose
evolution is governed by the equation
\begin{equation}
  \frac{dx(t)}{dt} = F(x;\Lambda = \{a,b,\mu\}) = 
  \left(a - b \, \mu \, \frac{\text{sign}
  \left(G(x)\right)}{2}\right)
  \vert G(x) \vert^{\mu -1} D(x),
  \label{eq1-det}
\end{equation}%
with initial condition $x(t_0) = x_0$.
Here, $a$, $b$ and $\mu$ are real numbers with
$b > 0$, $\text{sign}[\cdot]$ is the sign function,
$D(x)$ is a given everywhere non-negative function of $x$,
and $G(x)$ relates to $D(x)$ through
\begin{equation}
  \frac{dG(x)}{dx}=\frac{1}{D(x)}.
  \label{eq1a-det}
\end{equation}
Different $D(x)$'s specify distinct drift terms
and so we have in fact a set of systems.
Moreover, as we are going to see along this work, the present functional
form for our first order differential equation comply with
all the desired features listed in the beginning of the section.

\begin{figure}
\includegraphics[width=9.5cm]{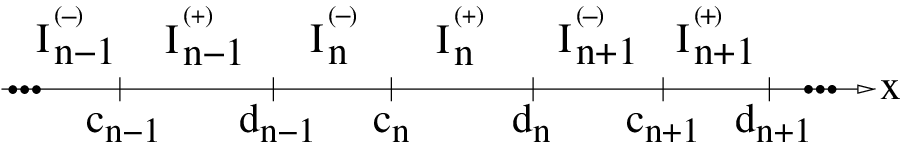}
\caption{\label{fig:fig1} Schematics of the successive distinct intervals along
  $x$ for which the function $G(x)$, basically the integral
  of $D(x)$, changes its sign.}
\end{figure}

Although $D(x) \geq 0$, $G(x)$ may assume positive or
negative values depending on $x$.
Thus, along the infinite line $x$ we can identify the intervals
$c_n < x < d_n$ as $I_n^{(+)}$ and $d_{n-1} < x < c_n$ as $I_n^{(-)}$
(see Fig. \ref{fig:fig1}), such that in $I_n^{(+)}$ ($I_n^{(-)}$)
the function $G(x) > 0$ ($G(x) < 0$).
By supposing $G(x)$ a continuous function, for any integer
$n \in (N_l,N_r)$, inevitably $G(c_n) = G(d_n) = 0$.
This is not the case if we allow ``jumps'' for $G(x)$
whenever $x$ crosses over between positive and negative intervals
$I^{(+)}$ and $I^{(-)}$.
If for all $x > x_r$ ($x < x_l$), $\text{sign}\left( G(x)\right) $ does not
change, then $N_r$ ($N_l$) is a finite integer, otherwise
$N_r \rightarrow \infty$ ($N_l \rightarrow - \infty$).

By noticing that $\vert G(x) \vert^{\mu -1} D(x)$ is never
negative, in principle the term $\text{sign}[G(x)]$ in Eq.
(\ref{eq1-det}) should give rise to a non-trivial spatially
asymmetric evolution depending on the parameter values.
To see why, let us consider that during the time interval
${\mathcal I}_T: t_0 \leq t < T$, the r-h-s of Eq.
(\ref{eq1-det}) does not vanish.
Hence
\begin{itemize}
\item{(i)}
For $|a|/b > |\mu|/2$ the time derivative of $x$ has
always the same sign of $a$.
As a consequence, $x(t)$ presents a steady increasing
($a > 0$) or decreasing ($a < 0$) behavior for $t \in
{\mathcal I}_T$ regardless of the functional form of
$G(x)$.
\item{(ii)} On the other hand, for $|\mu|/2 > |a|/b$, it follows
that $dx(t)/dt$ has the same sign of $- \mu \, G(x)$.
Therefore, along the evolution of $x(t)$ in the time interval
${\mathcal I}_T$, we have a switching in the variation of
$x(t)$ whenever $G(x(t))$ reverses its sign, conceivably
originating a rich dynamics depending on $D(x)$.
\end{itemize}

However, the system ceases to evolve when the r-h-s
of Eq. (\ref{eq1-det}) becomes zero.
This always takes place if along the time evolution of
$x(t)$ there is a $x(\overline{t}) = \overline{x}$ such that
either $a/b - \mbox{sign}[G(\overline{x})] \, \mu/2 = 0$
(in which case $|a|/b = |\mu|/2$) or 
\begin{equation}
  \vert G(\overline{x}) \vert^{\mu -1} D({\overline{x}}) = 0.
  \label{eq:condition}
\end{equation}

Disregarding the too specific and trivial situation of
$|a|/b = |\mu|/2$, for an oscillatory-like dynamics described
in (ii) to occur (say, within the interval $(x_{min},x_{max})$), Eq.
(\ref{eq:condition}) must be precluded.
For so, we observe that if $D(\overline{x}) = 0$
($\vert G(\overline{x}) \vert^{\mu -1} = 0$) then
$\vert G(\overline{x}) \vert^{\mu -1}$ ($D(\overline{x})$)
must diverge in such a way to maintain their product non-null
as $x \rightarrow \overline{x}$.
Therefore, we suppose that in the vicinity of $\overline{x}$ the
leading term from either a Taylor or a Laurent series for $D(x)$
and $G(x)$ is (for both $\nu, \gamma \neq 0$)
\begin{equation}
  D(x) \approx d \, (x - \overline{x})^\nu, \qquad
  G(x) \approx g \, (x - \overline{x})^\gamma,
\label{eq-series}
\end{equation}
with the constants $d, g \neq 0$.
So, we should have
\begin{equation}
 \nu + (\mu - 1) \, \gamma \leq 0 \ \ \  \Rightarrow \ \ \ 
 \nu \leq (1 -  \mu) \, \gamma.
 \label{exp-condition}
\end{equation}  
Moreover, from Eqs. (\ref{eq1a-det}) and (\ref{eq-series})
it follows that $g \, d \, \gamma \approx 1$ and
$\gamma + \nu \approx 1$, thus
\begin{equation}
  (2 - \mu) \, \nu \leq 1-\mu.
  \label{eq:mu-condition}
\end{equation}  
We remark that Eq. (\ref{eq:mu-condition}) fails when
$\mu=2$.

Particular cases are easily derived from the above.
Let us assume $\overline{x}$ in the interval of interest
$(x_{min},x_{max})$, then Eq. (\ref{eq:condition}) is not
verified at $x=\overline{x}$ in the following situations:
\begin{itemize}
\item{(a)} If $\mu = 1$, when $D(\overline{x}) \neq 0$.
  Thence, in the full $x$ interval we must have $D(x) > 0$.
\item{(b)} For $D(\overline{x}) = 0$ (so $\nu > 0$) then:
  (b-1) if $\mu < 1$, when $\nu \leq (1-\mu)/(2 - \mu) < 1$
  (so that $G(\overline{x}) = 0$);
  (b-2) if $\mu > 2$, when $\nu \geq (\mu - 1)/(\mu - 2) > 1$
  (such that $G(\overline{x})$ diverges).
\item{(c)} For $\mu > 1$ and $G(\overline{x}) = 0$, when
  $1 < \mu < 2$.
  In fact, here we must have a diverging
  $D(\overline{x})$, implying in $\nu < 0$.
  This condition requires the mentioned range for $\mu$.
\item{(d)} If there are jumps in $D(x)$, so that $G(x)$
  can change sign but without passing through zero,
  then when $D(x) > 0$.
  However, in such case we should define a proper prescription
  for Eq. (\ref{eq1a-det}) at these discontinuities.
\end{itemize}

It is clear from the above analysis that only for specifically
chosen functions $D(x)$ and parameter values, namely,
those observing the restrictions (a)--(d), the asymmetric term
in Eq. (\ref{eq1-det}) can yield a more diverse, eventually
stable or limited in space, dynamics.
This is in opposition to a simple, as sink or diverging, basins
of attraction emerging from Eq. (\ref{eq1-det})
for generic $D(x)$'s, i.e., functions not complying with
(a)--(d).

\begin{figure}
\includegraphics[width=9.5cm]{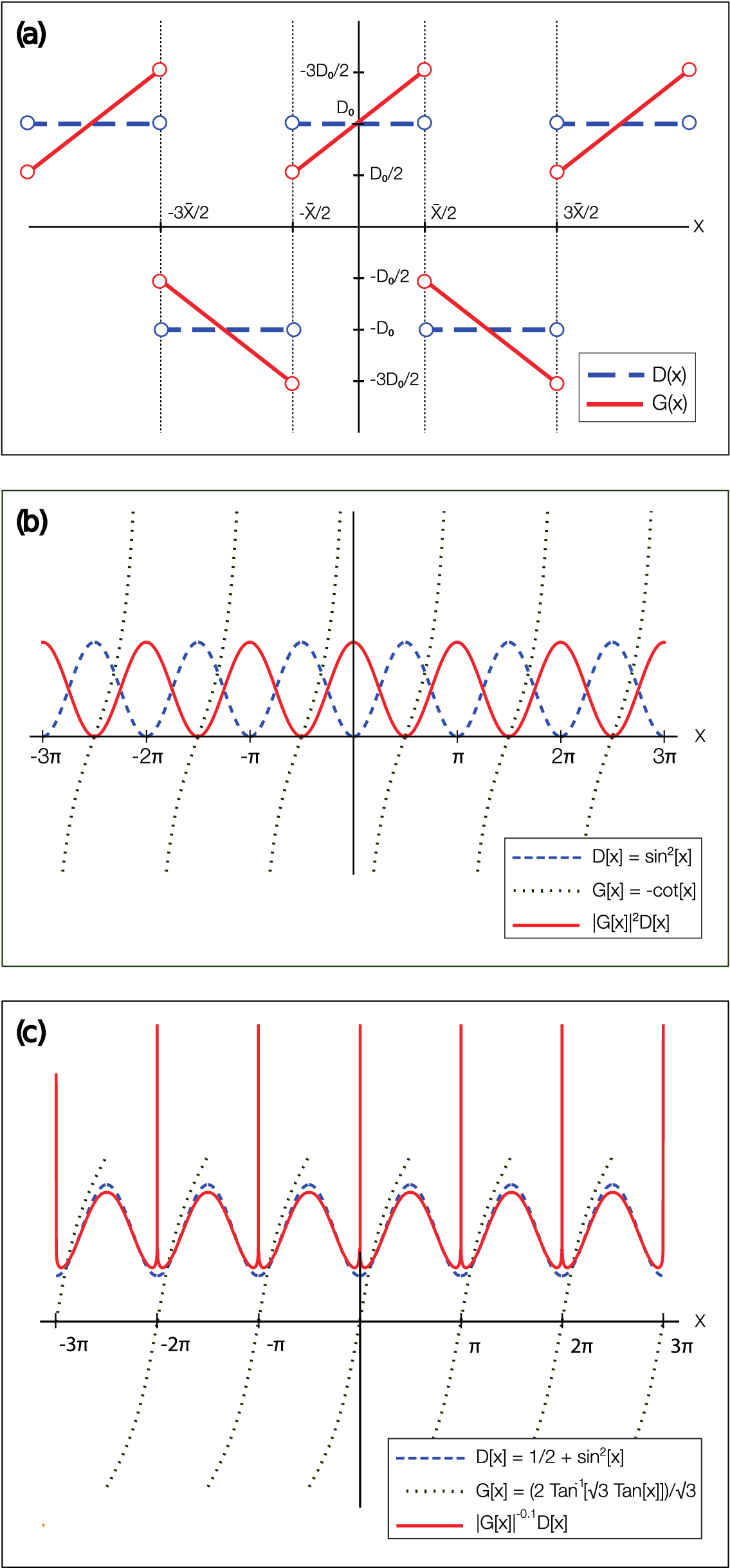}
\caption{\label{fig:fig2} Three examples of $D(x)$ and $G(x)$.
  In (a) and (c), the term $|G(x)|^{\mu-1} \, D(x)$ is never
  zero, potentially resulting in a rich dynamics for $x(t)$.
  But this requires very particular and singular functions.
  In (b) although $|G(x)|^{\mu-1} \, D(x)$ does not vanish
  at the points where $D(x) = 0$, this cannot be avoided
  for the points where $G(x) = 0$, eventually driving the
  system to a stationary behavior.}
\end{figure}

We now illustrate the previous discussion with three examples
of $D(x)$ and the associated $G(x)$ in Fig. \ref{fig:fig2},
assuming the parameters in the range specified by (ii).
In order to observe the condition (d),  we have functions with proper
jumps shown in Fig. \ref{fig:fig2} (a).
In this case $|G(x)|^{\mu-1} \, D(x)$ is never null, but at the
expense of rather specially tailored discontinuous $D(x)$ and
$G(x)$.
In Fig. \ref{fig:fig2} (b), $D(x) = \sin^2[x]$ and
$G(x) = -\mbox{cot}[x]$, with $\mu = 3$.
At the zeros of $D(x)$, one finds that $G(x)^2 \, D(x) > 0$,
hence precluding Eq. (\ref{eq:condition}), since the condition
(b) is verified (note that $G(x)$
diverges at these points).
Nevertheless, at the zeros of $G(x)$ we can not also avoid
$G(x)^2 \, D(x)$ to vanish, so with Eq. (\ref{eq:condition})
holding true.
Consequently, the system dynamics necessarily halts at these
sink points.
Lastly, for $D(x) = 1/2 + \sin^2[x]$,
$G(x) = 2 \, \mbox{artan}[\sqrt{3} \tan[x]]/\sqrt{3}$
and $\mu=11/10$, we have that $|G(x)|^{0.1} \, D(x)$ is never
null.
But this demands very narrow divergences for $|G(x)|^{0.1} \, D(x)$
(observe the spikes in the Fig. \ref{fig:fig2} (c)), hence
also for $dx(t)/dt$, whenever $x$ is very close to multiples
of $\pi$.
This type of drift might be unacceptable in modeling
distinct processes.

In this way, a natural question is: What would stabilize our
family of dynamical systems, averting a trivial evolution
from Eq. (\ref{eq1-det})  (where by trivial we mean $x(t)$
either becoming stationary or monotonically evolving towards
$\pm \infty$) for a much broader set of functions $D(x)$
and parameter values ?
We shall demonstrate below this can be achieved via
stochastic noise added to the original deterministic
problem.

For completeness, the formal general solution of the present
dynamical system in each interval $I^{(\pm)}$ is presented in the
 \ref{appendix1}.

\subsection{Adding stochastic noise}
\label{sec:model-stoc}

Hereafter we consider a noise term in our original family
of systems.
In doing so, we obtain a nonlinear Langevin equation
with space-dependent drift and diffusion coefficients and
driven by the Gaussian white noise (in the Stratonovich
description), or
\begin{equation}
  \frac{dx(t)}{dt} =
  \left(a - b \, \mu \, \frac{\text{sign}\left( G(x)\right) }{2}\right)
  \vert G(x) \vert^{\mu -1} D(x) +
  \sqrt{b} \, \vert G(x)\vert^{\mu /2} D(x) \, L(t) .
  \label{eq1}
\end{equation}%
The parameter ranges and relation between $D(x)$ and
$G(x)$ are as before.
The white noise force, $L(t)$, is such that \cite{risken}
\begin{equation}
  \left\langle L (t)\right\rangle = 0, \ \ \ \ 
  \left\langle L (t)L ({t}')\right\rangle =
  2 \, \delta (t-{t}'),
  \label{eqL2}
\end{equation}%
where $\delta (t)$ is the Dirac delta function.

Now we should emphasize that the
previous situation of dynamical traps for the deterministic model, 
represented by Eq. (\ref{eq:condition}), is far less common here.
Indeed, provided $a \pm b \, \mu/2 \neq 0$ and $\mu \neq 2$ in Eq.
(\ref{eq1}), for a given $x(t)=\overline{x}$ to lead to
$dx(t)/dt=0$ (so stationary) independently on the noise $L(t)$,
it should, at once, satisfy Eq. (\ref{eq:condition}) as well as
an akin relation with $\mu-1 \rightarrow \mu/2$ (for the
second term on the r-h-s of Eq. (\ref{eq1})).
But based on our previous analysis, this simultaneous condition
is very unlikely to happen for an arbitrary function $D(x)$.
Furthermore, for some very specific possible values of $L(t)$, 
instantly the sum of the two terms on the r-h-s of Eq. (\ref{eq1})
could become zero.
However, this exact cancelation would cease at subsequent times
as $L(t)$ varies.
  
The corresponding Fokker-Planck equation for the Langevin equation
(\ref{eq1}) in the Stratonovich approach is given by \cite{risken}
\begin{eqnarray}
   \frac{\partial \rho(x,t)}{\partial t}
  =-\frac{\partial }{\partial x} \! \!
  \left[\! \left( \! a \, \vert G(x)\vert^{\mu -1}+
    b \, \vert G(x)\vert^{\mu }\frac{dD(x)}{dx}\right) \! D(x)
    \rho (x,t)\right]+ \nonumber\\
b  \frac{\partial^2 }{\partial x^2} 
  \left[ \vert G(x)\vert^{\mu } D^2(x) \rho (x,t)\right] ,
 \label{eq2x}
\end{eqnarray}
where  $\rho\left( x,t\right)$ is the probability density function (PDF). 

\section{Exact Solutions and Analyses of Eq. (\ref{eq2x})}
\label{sec:gen-sol}

In this section we shall address the models described by Eq.
(\ref{eq2x}).
In doing so, we need to consider two distinct situations,
$\mu \neq 2$ and $\mu=2$.
Exact solutions are then derived by means of  variable
transformations and by using, respectively, the similarity method
for the former and the Fourier transform method for the latter.
More concrete and detailed examples are discussed in
Sec. \ref{sec:examples}.
Next, we will assume $G(\pm\infty)\to \pm\infty$.

\subsection{The case $\mu \ne 2$: Solutions from the
  similarity method}
\label{sec:sec3.1}

Equation (\ref{eq2x}) can be written as follows:
\begin{equation}
  \frac{\partial \rho (x,t)}{\partial t}
  =-a \, \frac{\partial }{\partial x}\Big[ \vert G(x)\vert^{\mu -1 }
    D(x) \, \rho (x,t) \Big] + b \, \frac{\partial }{\partial x}
  \left[ D(x) \frac{\partial }{\partial x}\big( \vert G(x)\vert^{\mu }
    D(x) \, \rho(x,t)\big) \right].
  \label{eq2}
\end{equation}%
Thus, from the transformations
\begin{equation}
  \bar{x}=G(x), \qquad \bar{\rho}
  (\bar{x},t) =D(x)\rho (x,t),
  \label{eq2ya}
\end{equation}%
Eq. (\ref{eq2}) reduces to
\begin{equation}
  \frac{\partial \bar{\rho}(\bar{x},t)}{\partial t}
  = -a \, \frac{\partial }{\partial \bar{x}}
  \Big[ \vert \bar{x}\vert^{\mu -1 } \, \bar{\rho}(\bar{x},t) \Big]
  + b \, \frac{\partial^2 }{\partial \bar{x}^2}
  \Big[ \vert \bar{x}\vert^{\mu } \, \bar{\rho}(\bar{x},t)\Big].
  \label{eq2z}
\end{equation}%
The above equation is invariant under the rescaling
(with $\gamma$ arbitrary)
\begin{equation}
  \bar{x} \to \epsilon \, \bar{x},~~t \to \epsilon^{2-\mu} \, t,
  ~~\bar \rho \to \epsilon^\gamma \, \bar\rho.
  \label{eq:renorm}
\end{equation}
Thus, we can employ the similarity solution method
\cite{kwoklivro,ho1,ho2} to address Eq. (\ref{eq2z}).

For $\bar{\rho}(\bar{x},t) = t^{-\alpha} \, \Phi(z)$ and
$z = \bar{x}/t^\alpha$, with the scaling exponent
$\alpha=1/(2-\mu)$, Eq. (\ref{eq2z}) reduces to the ordinary
differential equation (for ${\mathcal A}$ a constant)
\begin{equation}
  b \, \frac{d}{dz} \Big(|z|^\mu \, \Phi \Big) +
  \Big(\alpha z - a |z|^{\mu-1}\Big) \Phi = {\mathcal A}.
\label{eq3}
\end{equation}
By setting ${\mathcal A} = 0$ in Eq. (\ref{eq3})
(for our purposes here we do not need to address
the ${\mathcal A} \neq 0$ case), it readily follows that
\begin{equation}
  \Phi(z) = C \,
  \frac{|z|^{\frac{a}{b}\text{sign}(z)-\mu}}{b}
  \exp\left[-\frac{|z|^{2-\mu}}{b \, (2-\mu)^2}\right],
\label{Phi}
\end{equation}
where $C$ represents the normalization constant.
Therefore, a solution for Eq.\,(\ref{eq2x}) in the case of
$\mu\neq 2$ reads
\begin{equation}
  \rho \left( x,t\right) = C \,
  \frac{\vert G(x)\vert^{\frac{a}{b}\text{sign}\left( G(x)\right) -\mu}}
       {b \, D(x) \, t^{\frac{1-\mu +\frac{a}{b}\text{sign}\left( G(x)\right) }{2-\mu}}} \,
\exp\left[-\frac{\vert G(x)\vert^{2-\mu}}{b \, (2-\mu)^2 \, t}\right].
\label{eq1D3}
\end{equation}%
Note that the spatial asymmetry is manifested in Eq. (\ref{eq1D3})
through the term sign$[G(x)]$.
For a vanishing $a$, such asymmetry disappears.
We highlight that the PDF in Eq. (\ref{eq1D3}) can display
a broad range of behaviors depending on $D(x)$ and the corresponding
$G(x)$.

The normalization condition for $\rho(x,t)$ is the same
as that for $\Phi(z)$, so that we should have
\begin{equation}
  \int_{-\infty}^\infty \, \rho(x,t) \, dx
  = \int_{-\infty}^\infty \, \Phi(z) \, dz=1,
  \label{eq:integral}
\end{equation}
since $G(\pm \infty) \rightarrow \pm \infty$.
From Eq. (\ref{eq:integral}) we obtain the following normalization
constant by formal integration
\begin{equation}
  C = \frac{\left(b \, (2-\mu )^2\right)^{\frac{1}{2-\mu }}}
  {\vert 2-\mu \vert
    \left(\left( b \, (2-\mu )^2\right)^{-\frac{a}{b \, (2-\mu)}}
      \Gamma\left[\frac{1-\mu-\frac{a}{b}}{2-\mu}\right]
      +
      \left(b \, (2-\mu )^2\right)^{+\frac{a}{b \, (2-\mu )}}
      \Gamma\left[\frac{1-\mu+\frac{a}{b}}{2-\mu}\right]
      \right)},
  \label{eqCN2}
\end{equation}%
where $\Gamma[\cdot]$ denotes the Gamma function.
 We find that the PDF (\ref{eq1D3}) is not normalizable 
for $1\le\mu <2$ (actually, the similarity method is not the most
appropriate method to treat the $\mu=2$ case, see next section).
Similarly, one of the following two restrictions must also
be verified for a proper $C$:
\begin{itemize}
\item If $\mu < 1$,  \ \ \ then for a finite $C$ we further
  must have \ \ \ $\mu < 1 - |a|/b$,
\item If $\mu > 2$, \ \ \ then for a finite $C$ we further
  must have \ \ \  $\mu > 1 + |a|/b$. 
\end{itemize}

Usually, for an arbitrary $D(x)$ the computation of the $n$-moment,
given by $\langle x^n(t) \rangle$, is not an easy task.
On the other hand, the generalized $n$-moment
$\langle G^n(x) \rangle$ is far more amenable to calculations.
In order to obtain the generalized $n$-moment we take the whole
space (-$\infty$, $\infty$) for $x$, supposing a generic
$D(x) \ge 0$ (but we must notice that special cases might be simpler
to handle,
for instance, $G(x)=x$ if  $D(x)=1$ then we recover the ordinary
$n$-moment
$\langle x^n(t) \rangle = \langle G^n(x(t)) \rangle$).
Recall the extra condition,
$G(\pm \infty) \rightarrow \pm \infty$.
In this way, from
$\int_{-\infty}^\infty \, G^n(x) \, \rho(x,t) \, dx$ we have
 \begin{eqnarray}
  \langle G^n(x) \rangle =
  C \, \vert 2-\mu \vert
  \left(b \, (2-\mu )^2\right)^{\frac{n-1-\frac{a}{b}}{2-\mu}}
   \, t^{\frac{n}{2-\mu}} \times \nonumber\\
  \left((-1)^n \,
    \Gamma\left[\frac{n+1-\mu-\frac{a}{b}}{2-\mu} \right]
    + \left(b \, (2-\mu )^2\right)^{\frac{2\frac{a}{b}}{2-\mu}}
    \Gamma\left[\frac{n+1-\mu  +\frac{a}{b}}{2-\mu} \right]
    \right).
  \label{eq1D4}
\end{eqnarray}
Repeating the same type of analysis for 
Eq. (\ref{eq1D4}) as previously, we find that the generalized
$n$-moment is finite only if
\begin{itemize}
\item $\mu < 2:$ \ \ 
$n+1 > \mu + |a|/b$,
\item $\mu > 2:$ \ \ 
$n + 1 < \mu - |a|/b$.
\end{itemize}

\subsection{The case $\mu = 2$}

For the particular case of $\mu=2$ we employ variable
transformations and Fourier transform method. Now
we rewrite Eq. (\ref{eq2x})  as follows.
\begin{equation}
  \frac{\partial \rho (x,t)}{\partial t}
  =-\frac{\partial }{\partial x}\Big[ \left(a - b \,
    \text{sign}\left( G(x)\right) \right) H(x) \, \rho (x,t)\Big]
  + b \, \frac{\partial}{\partial x}
  \Big[H(x) \, \frac{\partial }{\partial x}
    \left(H(x) \, \rho (x,t) \right) \Big],
  \label{eqSol2a}
\end{equation}
where
$H(x) = \vert G(x)\vert \, D(x)$.
For simplicity, we restrict ourselves to the case where
$\text{sign}(G(x))$ assumes a unique value.
Considering
\begin{equation}
  \frac{dx^{*}}{dx} = \frac{1}{H(x)}
  \ \ \text{and} \  \ \rho^{*} (x,t) = H(x) \, \rho(x,t),
  \label{eq2y}
\end{equation}%
we obtain the following Fokker-Planck equation:
\begin{equation}
  \frac{\partial \rho^{*} (x^{*},t)}{\partial t}
  = -
  \frac{\partial}{\partial x^{*}}
  \big[ \left( a - b \, \text{sign}\left( G(x)\right)
    \right) \rho^{*} (x^{*},t)
  \big] + b \,
    \frac{\partial^2}{\partial x^{*2}} \rho^{*}(x^{*},t) ;
 \label{eqSol2b}
\end{equation}%
its solution is obtained from the Fourier transform method supposing
the initial condition $\rho^{*} (x^{*},0)=\delta (x^{*}-x_0^{*})$.
The final result is then \cite{kwok5}
\begin{equation}
  \rho \left(x,t\right) =
  \frac{C}{\sqrt{4 \pi \, b \, t} \, H(x)}
  \exp\left[-\frac{(x^*(x)-x_0^*- (a - b \,
      \text{sign}\left( G(x)\right) ) \,
      t)^2}{4 \, b \, t}\right],  
\label{eqSol2c}
\end{equation}%
where $C$ is the  normalization constant.

We should remark that the PDF in Eq. (\ref{eqSol2c}) is also
the solution for the case of zero drift, but for the position
coordinate $x^{*}$ translated by
\begin{equation}
  x^*(x) \rightarrow x^*(x) -
  (a-b \, \text{sign}\left( G(x)\right) ) \, t.
  \label{eqSol2d}
\end{equation}%

\subsection{The qualitative dynamical evolution of
  the deterministic and stochastic models in the
  parameters space}

\begin{figure}
\includegraphics[width=15cm]{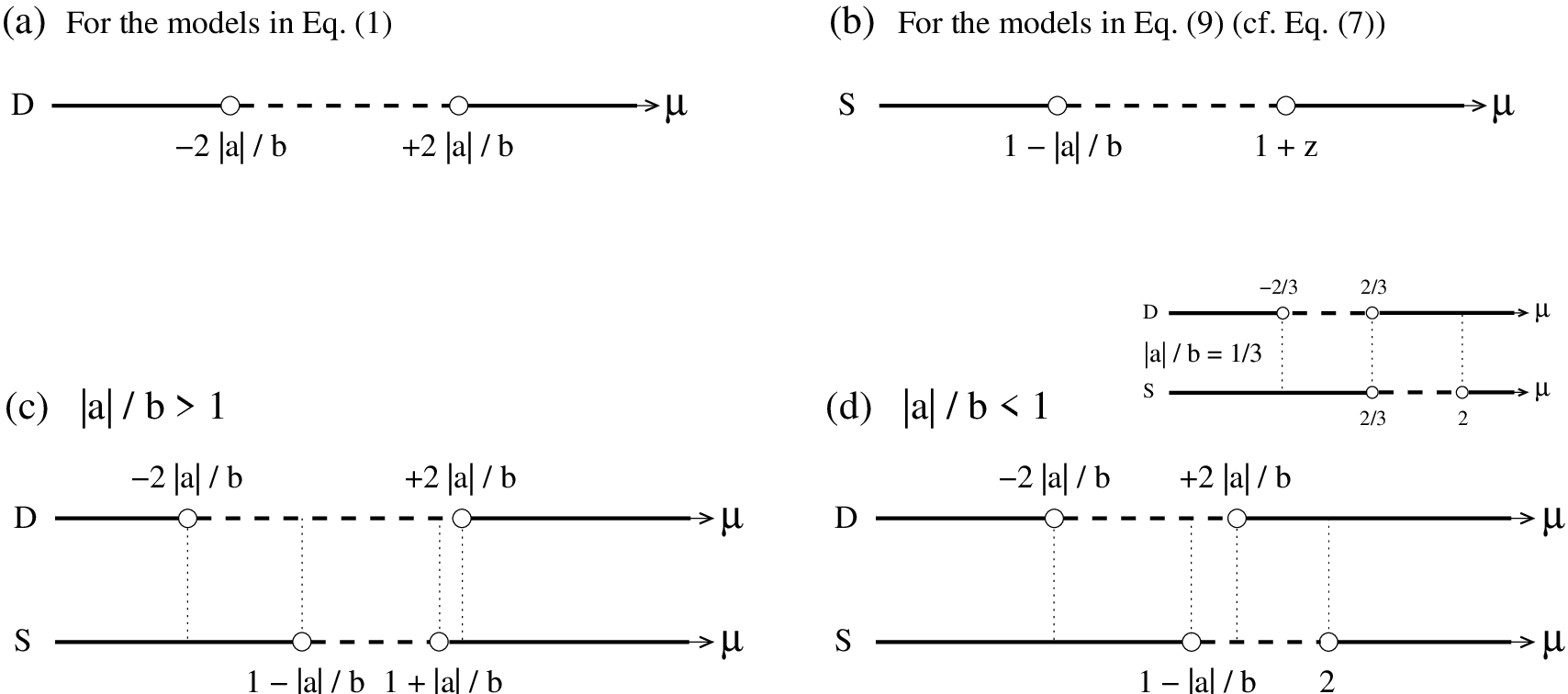}
\caption{\label{fig:fignew} (a) In the absence of dynamical traps
  (see Sec. \ref{sec:model-det}), for the deterministic models D
  the $\mu$ parameter intervals for which the trajectories
  either steadily tend to infinity (dashed segment, of length
  $\Delta \mu_{{\mbox{\scriptsize D}}} = 4 \, |a|/b$) or have a richer
  oscillating behavior, confined to a certain $x$ spatial region
  (continuous semi-lines).
  (b) For the stochastic models S, the continuous semi-lines
  (dashed segment, of length $\Delta \mu_{{\mbox{\scriptsize S}}} =
  z + |a|/b$) indicate the $\mu$ parameter values for which
  $\rho(x,t)$ has (has not) well-behaved analytic expressions.
  Here $z = \mbox{max}(1,|a|/b)$.
  Defining $r =
  \Delta \mu_{{\mbox{\scriptsize D}}}/\Delta \mu_{{\mbox{\scriptsize S}}}$,
  the $\mu$ interval for D leading to diverging
  trajectories is always longer than those for S presenting
  no proper $\rho(x,t)$ solutions provided $|a|/b > 1/3$.
  Indeed, (c) $r = 2$ for $|a|/b > 1$ and (d)
  $r = 4 \, (|a|/b)/(1 + |a|/b) > 1$ if $1>|a|/b > 1/3$
  (the inset depicts the $|a|/b=1/3$ case). }
\end{figure}

In the previous sections we have analyzed the features of the
deterministic and stochastic models in terms of ranges of
values for the parameters.
In particular, we have unveiled that in most instances, the
non-linear deterministic models of Eq. (\ref{eq1-det}) display
rather straightforward (monotonous) trends.
In fact, they would require specific conditions ---
both for the set $\Lambda = \{a, b, \mu\}$ as well for the properties
of $D(x)$ and $G(x)$ --- so to yield a more diverse
dynamics for $x(t)$, say oscillating between two extrema,
$x_{min}$ and $x_{man}$, instead of approaching a fixed point
$\overline{x}$ in finite time or asymptotically evolving towards
$\pm \infty$.

Conversely, by adding white noise to our original problem, we have
obtained a non-linear Langevin equation, given by Eq. (\ref{eq1}).
In this case dynamical traps are far more rare and the evolution
of the now stochastic $x(t)$ is not affected by most of the
restrictions discussed for the deterministic case in
Sec. \ref{sec:model-det}.
However, rather than addressing in detail such stochastic
microscopic variable, we have followed the standard procedure of
considering the associated probability density function $\rho(x,t)$,
seeking for general non-trivial solutions displaying spatial
asymmetry.
For instance, notice that relevant changes of behavior for
$\rho$ arise depending on the term sign$(G(x))$ appearing
in Eq. (\ref{eq1D3}) (explicit examples in the next section).

Thus, keeping in mind the obvious conceptual differences between
the physical meaning of $x(t)$ governed by Eq. (\ref{eq1-det})
and $\rho(x,t)$ by Eq. (\ref{eq2}), one can qualitatively contrast
their dynamics considering the ranges of $\Lambda =
\{a, b, \mu\}$ (see Fig. \ref{fig:fignew} (a) and (b)),
which in one hand may result in simple monotonic evolution for
the deterministic systems, but on the other hand might allow
phenomenologically much more diverse behavior for the stochastic
models (cf., Eq. (\ref{eq1D3})).
This kind of comparison can be viewed as a heuristic (but not
sensibly departing from more rigorous approaches in the
literature, e.g., as those in \cite{raducha-2020,bellavia-2023})
or even an operational way of inferring how stochasticity can lead
to the emergence of complexity \cite{lindner-2004} in certain
classes of processes such as \cite{agazzi-2002,albeverio-2022}:
dynamics in networks, lasing in noisy media, pattern-formation,
granular matter nucleation and ecological interactions, to cite
a few examples.

So we need to address only the instances where $x(t)$
in Eq. (\ref{eq1-det}) can asymptotically diverge since the
scarcity of dynamical traps for Eq. (\ref{eq1}) has already
been discussed (see the paragraph following Eq. (\ref{eqL2})
in Sec. \ref{sec:model-stoc}).
We restrict the analysis to $\mu \neq 2$.
For our purposes we rewrite Eq. (\ref{eq1}) as
\begin{equation}
\frac{dx(t)}{dt} = {\mathcal F}(x) \,
\Big({\mathcal G}(x) + L(t)/\sqrt{b}\Big),
\label{eq:teste}
\end{equation}
where ${\mathcal F}(x) = b \, |G(x)|^{\mu/2} \, D(x) \geq 0$
and ${\mathcal G}(x) = (a/b - (\mu/2) \, \mbox{sign}(G(x))) \,
|G(x)|^{\mu/2 - 1}$.
We also recall the condition (i) in Sec. \ref{sec:model-det},
namely,
\begin{equation}
  |a|/b > |\mu|/2.
  \label{eq:cond1}
\end{equation}
For it (cf. Fig. \ref{fig:fignew} (a)), regardless of $G(x)$ the
dynamical evolution of the deterministic models in the absence
of traps are monotonic, i.e., $x(t)$ invariably just increases or
just decreases with $t$.
Moreover, by inspecting ${\mathcal G}(x)$ in Eq. (\ref{eq:teste})
and considering the relation in Eq. (\ref{eq:cond1}) it follows
that ${\mathcal G}(x)$ is either always positive ($+$) or always
negative ($-$).
So, in the long run $x(t)$ should not diverge (i.e., in average
not evolving towards $\pm \infty$) only if the fluctuations
from $L(t)/\sqrt{b}$ could refrain this biased evolution of $x(t)$
driven by ${\mathcal G}(x)$.
Observe that in such a context, $x(t)$ does resemble a
random walk, but with a spatial bias in a given direction.
In other words, in this situation --- which we call the
non-stabilization condition --- the added noise cannot confine
the systems.
We shall emphasize the known fact \cite{bonet-1995}
that for a non-linear Langevin equation, certain characteristics
of the resulting non-linear trajectories, noticeably divergence
\cite{ryabov-2019,mazumdar-2020}, may hinder a proper PDF
description via a linear Fokker-Planck equation (for
a comprehensive discussion see, e.g., \cite{callaham-2021}).
Particularly, it poses important issues related to stability
and solvability of the latter \cite{mazumdar-2020,zhao-2022}.

Then, first consider $|a|/b > 1$.
When $\mu > 2$ ($\mu < 1$), from Eq. (\ref{eq:cond1}) the
deterministic models are monotonic for $2 < \mu < 2 \, |a|/b$
($-2 \, |a|/b < \mu < 1$), tending to $\pm \infty$ if there are
no dynamical traps along the way.
But when $\mu > 2$ ($\mu < 1$), the stochastic models are
well-behaved for $\mu > 1 + |a|/b$ ($\mu < 1 - |a|/b$),
Sec. \ref{sec:sec3.1}.
Hence, stabilization through the addition of white noise is
attained in the ``extra'' intervals 
$1 + |a|/b < \mu \leq 2 \, |a|/b$ and
$-2 |a|/b \leq \mu < 1 - |a|/b$, Fig. \ref{fig:fignew} (c),
representing a considerable range increasing  along $\mu$ of $\Delta\mu=2 \, |a|/b$.
In the remaining interval $1 - |a|/b < \mu < 1 + |a|/b$ for the
stochastic models --- in which Eq. (\ref{eq1-det}) also leads to
diverging trajectories --- the function ${\mathcal G}(x)$ in
Eq. (\ref{eq:teste}) has always the same sign and the noise 
is not enough to avoid the asymptotic natural leaning.
Consequently, for $1 - |a|/b < \mu < 1 + |a|/b$ the mentioned
non-stabilization condition applies.

Second, assume $|a|/b < 1$.
The distinction is that now the stochastic models have no solutions
for $1 - |a|/b < \mu < 2$ (instead of $1 - |a|/b < \mu < 1 + |a|/b$).
In this way, the analysis for the left limits are akin to
$\mu < 1$ above, compare Fig. \ref{fig:fignew} (c) and (d).
Thus, we can focus only on the right limits.
Observe that in the range $1 - |a|/b < \mu < 2 \, |a|/b$, Fig.
\ref{fig:fignew} (d), the previous non-stabilization condition
takes place.
On the contrary, although the deterministic systems are not
diverging in the interval $2 \, |a|/b < \mu < 2$, the stochastic
ones have no solutions.
For the time being we have not found a more conceptual explanation
--- mathematically, they are those in Sec.  \ref{sec:sec3.1} ---
for such result (hopefully, it will be elucidated in a
forthcoming contribution).
But the point is that by the inclusion of noise, the interval
of diverging trajectories for the deterministic models,
$4 \, |a|/b$, is larger than the interval of non-normalizable
solutions for the stochastic models, $1 + |a|/b$, whenever
$|a|/b > 1/3$.
We remark that $|a|/b = 1/3$ is the threshold to exist
an overlap between the intervals $(-2 \, |a|/b, +2 \, |a|/b$)
and $(1-|a|/b, 2)$, respectively, for the deterministic and
stochastic models, see the inset of Fig. \ref{fig:fignew} (d).
The qualitative reason for this borderline $|a|/b = 1/3$ value
also requires future investigations.

\section{Some specific examples for the stochastic model}
\label{sec:examples}

Next we illustrate by means of simple, but representative,
examples some trends of the PDF $\rho$'s presented in Sec.
\ref{sec:gen-sol}.
In Sec. \ref{sec:examples-1} we consider $\mu \neq 2$
and $D(x) = 1$.
This is an interesting choice because in this case the
generalized $n$-moment reduces to the standard one (
$\langle x^n \rangle$).
In Sec. \ref{sec:example-mu2} we address $\mu = 2$.
In particular, for a specific $D(x)$ we show that our
system relates to important population growth models
in the literature.

\subsection{The case of $\mu \neq 2$ and $D(x)=1$}
\label{sec:examples-1}

For $D(x) = 1$ we have $G(x) = x \,  + \, $constant.
For simplicity we set such constant to zero.
We just comment that depending on $a$, $b$ and $\mu$ (and the
initial condition $x_0$), the dynamical system can either become
stationary or diverging.

In Eq. (\ref{eq1}), $|G(x)|^{\mu-1} = |x|^{\mu-1}$ and
$|G(x)|^{\mu/2} = |x|^{\mu/2}$ give rise to distinct power-law
functions.
The normalized PDF reads (for $C$ in Eq. (\ref{eqCN2}))
\begin{equation}
  \rho(x,t) = C \,
  \frac{\vert x\vert^{\frac{a}{b}\text{sign}(x)-\mu}}{b \,
    t^{\frac{1-\mu +\frac{a}{b}\text{sign}(x)}{2-\mu}} }\,
\exp\left[-\frac{\vert x\vert^{2-\mu}}{b \, (2-\mu)^2 \, t}\right].
  \label{eqPDF2}
\end{equation}
It is interesting to note that the PDF given by Eq.
(\ref{eqPDF2}),
for $\mu=a/b$ and $\mu <1$, is composed  of a stretched or
compressed Gaussian distribution and a generalized Weibull
distribution.
This dual functional form suggests that the system described by
Eq. (\ref{eq1}) may be used to model processes resulting from
different dynamical drives.

The $n$-moment can be obtained from Eq. (\ref{eq1D4}), yielding
\begin{eqnarray}
  \left< x^n \right> &=& C \,
  \vert 2-\mu \vert \left(b \,
  (2-\mu )^2\right)^{\frac{n-1-\frac{a}{b}}{2-\mu}}
  \left( (-1)^n\, \Gamma\left[\frac{n+1-\mu  -
      \frac{a}{b}}{2-\mu} \right] \right.
  \nonumber \\
  & & + \left. \left(b \, (2-\mu )^2\right)^{\frac{2\frac{a}{b}}{2-\mu}} \,
\Gamma \left[
\frac{n+1-\mu  +\frac{a}{b}}{2-\mu} \right] \right) \,
t^{\frac{n}{2-\mu}}
\label{eqAS3}
\end{eqnarray}  

In the particular case of $a=0$, the PDF in Eq. (\ref{eqPDF2})
reduces to 
\begin{equation}
  \rho(x,t) =
  \frac{\left(b \, (2-\mu )^2\right)^{\frac{1}{2-\mu }}\vert x\vert^{-\mu}}
       {2\, b \, \vert 2-\mu \vert  \, \Gamma
         \left[\frac{1-\mu}{2-\mu }\right] \,
         t^{\frac{1-\mu}{2-\mu}}} \,
       \exp\left[-\frac{\vert x\vert^{2-\mu}}{b(2-\mu)^2t}\right],
\label{eqPDF3}
\end{equation}%
and Eq. (\ref{eqAS3}) results in (with $C_0 = C|_{a=0}$)
\begin{equation}
  \left< x^n\right> = C_0 \,
  \vert 2-\mu \vert \left(b \, (2-\mu )^2\right)^{\frac{n-1}{2-\mu}}
  \, \Gamma \left[\frac{n+1-\mu  }{2-\mu} \right]
((-1)^n+ 1) \, t^{\frac{n}{2-\mu}} . \label{eqAS1D4}
\end{equation}%
One can see that $\left<x^n\right>$ in Eq. (\ref{eqAS1D4}) is zero
for $n$ an odd number, in accordance with the symmetric PDF in 
Eq. (\ref{eqPDF3}).
In particular, its second moment goes with $t^{2/(2-\mu)}$, hence it
can describe superdiffusive, normal and subdiffusive, processes
respectively for, $0< \mu < 1$, $\mu =0$ and $\mu < 0$.
Further, for $\mu >2$ the system describes localized processes.

Generally, the PDF in Eq. (\ref{eqPDF2}) represents a system
with a power-law potential of order higher than $2$ when $\mu > 2$
and with a spatial asymmetry associated with the
term $a \, \text{sign}(x)$, or
(with $a\ne 0$ and $ b \, \mu/2- a {\rm\ sign} (x) > 0$)
\begin{equation}
  V(x) \sim \frac{(b \, \mu/2 - a \ \text{sign}(x))}
  {\mu} \,  \vert x\vert^{\mu}.
  \label{eqCoe21}
\end{equation}%
Thus, the parameter $|a|$ determines the degree of asymmetry, with
$a=0$ leading to a totally symmetric PDF, Eq. (\ref{eqPDF3}).

The solution given by Eq. (\ref{eqPDF2}) is not valid for $\mu=2$.
Nonetheless, we can take $\mu \sim 2$, so that
$|G(x)|^{\mu-1}$ (related to the non-linear drift and with the
asymmetry term sign$(G(x))$) and
$|G(x)|^{\mu/2}$ (related to the white noise coefficient)
in Eq. (\ref{eq1}) are both approximately linear in $|G(x)|$.
In this case the $V(x)$ in Eq. (\ref{eqCoe21}) with $\mu \sim 2$
fairly represents the usual symmetric (asymmetric) harmonic
potential for $a=0$ ($a\ne 0$ and $b/2>\vert a\vert$).

We  can also compare the PDF in Eq. (\ref{eqPDF3}) with that
one obtained from Eq. (\ref{eq1}), but without drift and derived
from different prescriptions, or (see the ref. \cite{kwoklivro})
\begin{equation}
  \rho(x,t) =
  \frac{\vert x\vert^{-(1-\lambda ) \, \mu}}
       {2 \, \vert 2-\mu \vert^{ \frac{-(1-2\lambda) \, \mu}{2-\mu }} \,
         \Gamma\left[\frac{1-(1-\lambda)\mu}{2-\mu }\right]
         (b\, t)^{\frac{1-(1-\lambda) \, \mu}{2-\mu}}} \,
\exp\left[-\frac{\vert x\vert^{2-\mu}}{b \, (2-\mu)^2 \, t}\right].
       \label{eqPDF4}
\end{equation}%
Here $0 \le \lambda \le 1$ is the prescription parameter,
for $\lambda=1/2$ yielding the Stratonovich's and
$\lambda=0$ the Ito's.
The $n$-moment related to the PDF (\ref{eqPDF4}) is given by
\begin{equation}
  \left< x^n(x)\right> =\frac{ \left( b \, (2-\mu )^2\right)^{\frac{n}{2-\mu}}
    \, \Gamma \left[\frac{n+1-(1-\lambda ) \, \mu}{2-\mu} \right]
    \, t^{\frac{n}{2-\mu}}}{\Gamma \left[\frac{1-(1-\lambda ) \, \mu}{2-\mu}
      \right]},
  \label{eqPDF5}
\end{equation}%
where $n$ is an even number.
Observe that Eq. (\ref{eqPDF4}) coincides with Eq. (\ref{eqPDF3})
for $\lambda = 0$ (the Ito prescription).
This follows directly from the fact that for the Fokker-Planck
equation in Eq. (\ref{eq2x}), the drift term vanishes since for
$D(x)=1$ we have $dD(x)/dx=0$.
Moreover, the $n$-moment in Eq. (\ref{eqPDF5}) has similar behavior
to that one given by Eq. (\ref{eqAS1D4}) for even numbers.

All these findings show that there are a large class of systems
displaying the same $n$-moment trends.

\begin{figure}
\includegraphics[width=0.7\linewidth]{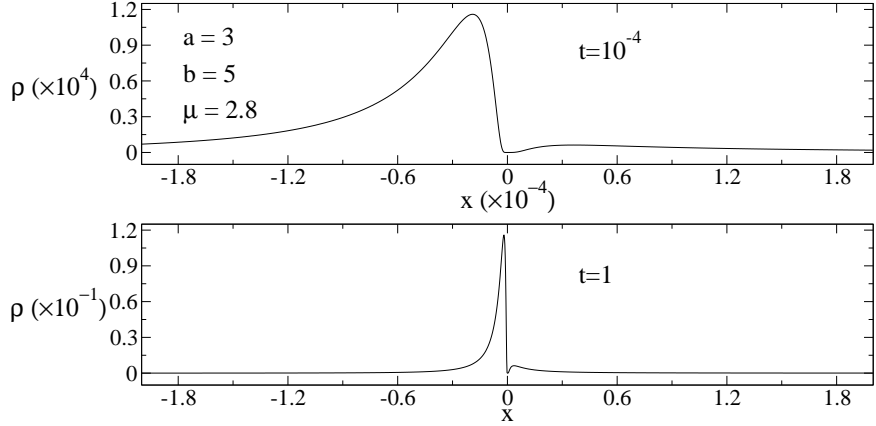}
\caption{For the case of $D(x) = 1$, the
  resulting PDF, Eq.
  (\ref{eqPDF2}), at two time instants, $t=1$ and the very
  short $t=10^{-4}$ (on purpose, to illustrate the shape
  of the initial PDF).
  For the parameters $a$, $b$ and $\mu$ considered,
  $\mu > 2 > 1 + a/b > 2 \, a/b$, thus not belonging to the
  monotonic behavior (i) for the dynamical system.
  Although the distribution is considerably higher for $x < 0$,
  it is still noticeable for $x$ positive, but only in the origin
  vicinity. In the plots, both $\rho$ and $x$ have been rescaled so
  to facilitate a direct comparison between the curves shapes.
  Notice that $\rho$ is spatially much more concentrated at
  $t=1$ than at $t=10^{-4}$.}
\label{fig:fig3}
\end{figure}

\begin{figure}
\includegraphics[width=0.7\linewidth]{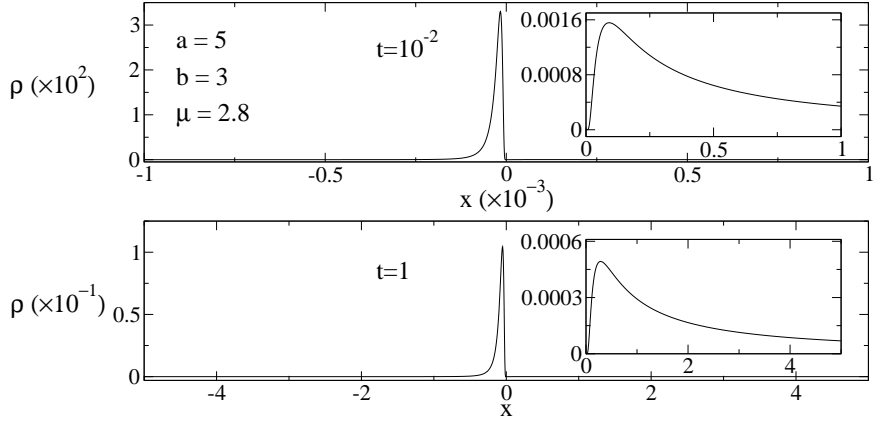}
\caption{Similar plots as in Fig. \ref{fig:fig3}, but for the
  parameters values such that $2 \, a/b > \mu > 1 + a/b > 2$.
  Hence, the corresponding dynamical system does belong to the
  monotonic behavior class (i).
  Since now the PDFs are very small for $x$ positive, the insets
  show proper blow ups for $x > 0$.
  Again, $\rho$ is spatially much more concentrated at
  $t=1$ than at $t=10^{-2}$.
  }
\label{fig:fig4}
\end{figure}

\begin{figure}
\includegraphics[width=0.7\linewidth]{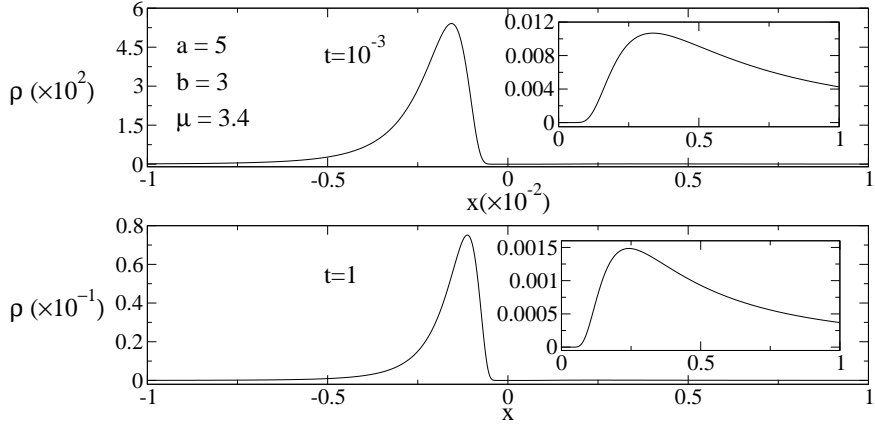}
\caption{
 Similar plots as in Fig. \ref{fig:fig4}, but for the
  parameters satisfying $\mu > 2 \, a/b > 1 + a/b > 2$.
  Thus, as in Fig. \ref{fig:fig3}, not belonging to the monotonic
  behavior (i) for the dynamical system.
  The PDF is spatially much more concentrated at
  $t=1$ than at $t=10^{-3}$.}
\label{fig:fig5}
\end{figure}

\begin{figure}
\includegraphics[width=0.7\linewidth]{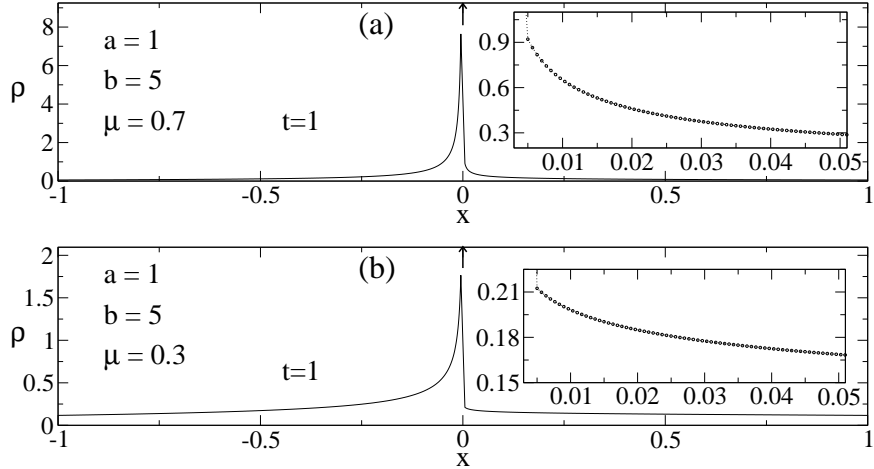}
\caption{For $D(x) = 1$, the PDF in Eq.
  (\ref{eqPDF2}) at $t=1$ with $a=1$, $b=5$ and two
  values of $\mu < 1$.
  (a) $\mu = 0.7$, thus  $2 \, a/b < \mu <  1 - a/b < 1$.
  (b) $\mu = 0.3$, thus  $\mu < 2 \, a/b <  1 - a/b < 1$.
  The arrows indicate that although integrable, these
  $\rho$'s tend to $+\infty$ at the origin.
  The insets give details of the distributions for $x>0$
  very close to the origin.
  These PDF's tend to be more concentrate as $t$ increases,
  plots not shown.}
\label{fig:fig6}
\end{figure}

Finally, graphs of Eq. (\ref{eqPDF2}) for distinct parameter
values (all with $a > 0$) and at different time instants $t$
are depicted in Figs. \ref{fig:fig3}--\ref{fig:fig6}.
For $a \rightarrow -a$ we get a specular image, about $x=0$, of the
observed profiles.
Since $G(x) = x$,  which is anti-symmetric in $x$, all the plots
display an imbalance regarding positive and negative $x$'s, hence
overall with $\rho(x<0)$ much greater than $\rho(x>0)$.
Also, the imbalance tends to be stronger for greater $|a|$'s.
For Figs. \ref{fig:fig3}--\ref{fig:fig5} (\ref{fig:fig6}
(a) and \ref{fig:fig6} (b)) we have $\xi = \mu - 1 > 0$
($\xi = 1 - \mu > 0$), so that $|G(x)|^{\mu-1} \, D(x) = |x|^\xi$
($|G(x)|^{\mu-1} \, D(x) = 1/|x|^\xi$).
This explains why the corresponding PDFs are very small (very large)
for $x$ approaching zero.
From the plots for $\mu > 2$ we see that as $t$ increases, the
distributions tend to concentrate around the origin.
We have checked this is likewise the case for the examples with
$\mu < 1$ (not shown), but then with such concentration taking
place slower in time.
As a last remark, provided $a$ and $b$ are the same and 
the values of the associated $\mu$'s do not differ much,
 we have not detected relevant qualitatively
differences among the $\rho$'s either when
$2 \, a/b <  \mu < 1 - a/b$ or when
$\mu < 2 \, a/b <  1 - a/b$ if $\mu < 1$ and when
$\mu > 1 + a/b > 2\, a/b$ or when $2 \, a/b > \mu >  1 + a/b$
if $\mu > 2$.

\subsection{Some examples for $\mu=2$}
\label{sec:example-mu2}

We finally consider the solution in Eq. (\ref{eqSol2c}).
For $H(x)=\sqrt{h}$, with $h$ a positive constant, we have
$x^{*}(x)=x/\sqrt{h}$, $\vert G(x)\vert = \exp[\pm x/\sqrt{h}]$
and $D(x) = \sqrt{h} \ \exp[\mp x/\sqrt{h}]$.
In this case, the PDF in Eq. (\ref{eqSol2c}) recovers the well-known
$\rho$ for a Brownian motion with a load force, whose expression
is \cite{kwokload}
\begin{equation}
  \rho(x,t) = \frac{1}{\sqrt{4 \, \pi \, b \, h \, t}}
  \exp\left[-\frac{\big(x-x_0-\sqrt{h} \, (a-b) \, t\big) ^2}
    {4 \, b \, h \, t}\right].
  \label{eqMod2a0}
\end{equation}%


A particularly interesting situation, relating to other models
in the literature, results from $H(x)$ given by
\begin{equation}
  H(x) =
  \frac{\left( K^{\alpha }-x^{\alpha }\right) \, x}
       {\beta \, K^{\alpha } - \left(\beta -\alpha\right)
         \, x^{\alpha }},
\label{eqMod2a}
\end{equation}%
such that
\begin{equation}
  D(x) =
   \frac{K^\beta}{K^\alpha} \,
  \frac{\left(K^{\alpha }-x^{\alpha }\right)^2
    \, x^{1-\beta}}{(\beta \,
    K^{\alpha }-\left(\beta -\alpha\right) \, x^{\alpha })}, \ \ \ 
  G(x)=  \frac{\left(x/K\right) ^{\beta }}
  {1-\left(x/K\right) ^{\alpha}}.
\label{eqAp24}
\end{equation}%
In this case, the PDF in Eq. (\ref{eqSol2c}) is related to the
population growth model proposed in \cite{saka}
(see also \cite{kwok3}).
More specifically, $0 \leq x(t) \leq K$ is the number of alive
individuals in a population at time $t$,
$r=a-b \, \text{sign}\left(G(x)\right) = a - b$ is the intrinsic
growth rate with $a>b$, and $K$ is the carrying capacity.
For simplicity, the parameters $\beta $ and $\alpha $ are
restricted to real non-negative values.
In fact, the system described by Eqs. (\ref{eq1}) and (\ref{eqMod2a})
encompasses classical growth models such as the
Verhulst logistic ($\beta =1$ and $\alpha =1$),
Gompertz ($\beta =0$ and $\alpha \rightarrow 0$),
Shoener ($\beta =0$ and $\alpha =1$),
Richards ($\beta =0$ and $0< \alpha <\infty $)
and Smith ($0\leq \beta <\infty $ and $\alpha =1$)
\cite{calisto2,popul,kwok3,richa,saka,goel,roman,gomp}.
Also, for $\beta = \alpha =1$ the present framework has been
employed in the study of population genetics \cite{kimura}.

In formulating the above mentioned models through the present
approach, some care is necessary concerning the normalization constant.
For instance, to avoid a time dependent $C$ (implying in a
non-conservation of probability along $t$), the limits of
integration for $\rho$, $x_i^*(x=0)$ and $x_f^*(x=K)$, should
not be finite.
This is determined from (recall that $dx^*/dx = 1/H(x)$)
\begin{equation}
  x^{*} = \ln\left[\frac{\left(x/K\right)^{\beta}}
    {1-\left(x/K\right)^{\alpha}}\right].
\label{eqAp21}
\end{equation}%
As an example we consider the Shoener and Richards models
(both with $\beta = 0$).
They cannot be constructed under the present procedure once the
lower limits have finite values.
Indeed, for $\beta = 0$ and $\alpha > 0$ we get
$x^*=-\ln[|1-\left( x/K\right) ^{\alpha}|]$,
so that $x_i^*$ is zero for $x=0$. 

For $\beta, \alpha \neq 0$ we have
$x^{*}(x\rightarrow 0)\rightarrow -\infty$ and
$x^{*}(x\rightarrow K)\rightarrow \infty $.
Thus, from Eq. (\ref{eqSol2c}) we find $C=1$ and the
PDF yields (with $t_0 = 0 $)
\begin{equation}
  \rho(x,t) =
  \frac{1}{\sqrt{4 \, \pi \, b \, t} \, H(x)}
  \, \exp\left[-\frac{\big(x^*(x)-x_0^*-\left( a-b\right) t \big)^2}
    {4 \, b \, t}\right].
 \label{eqAp25}
\end{equation}%
In general $x^{*}_0$ is considered finite (except for
$\alpha \rightarrow 0$).
Thus, one should exclude a population that is initially null
($x_0 \rightarrow 0$) or that is already at its maximum possible
value established by $K$ ($x_0\rightarrow K$)

\section{Conclusion}

In this contribution we have considered a set of models with
asymmetric space-dependent drifts.
For the deterministic case, we have identified their temporal
evolution features in the parameters space $\Lambda = \{a, b, \mu\}$
and also in terms of certain general properties of the driven
function $D(x)$ (and of its primitive integral $G(x)$).
We have shown the deterministic models display rather simple
behavior in a large region of $\Lambda$.

Then, by adding the Gaussian white noise to such class of problems
we have obtained nonlinear Langevin equations, whose associated
Fokker-Planck equations have been solved through the similarity
method or Fourier transform method.
In the subset of $\Lambda$ where the obtained $\rho$'s are well
behaved, we have discussed the dynamical richness emerging
from these PDFs.
For instance, conceivably they could be used to study anomalous
diffusion, with applications in different processes as
population growth models.

By comparing the two families of models in $\Lambda$, we have
unveiled the effects of introducing stochasticity and the
mechanisms allowing the qualitative changes observed in the
systems dynamics.
Concretely, we have found that for some regions of $\Lambda$,
although the trajectories of the deterministic models are trivial,
i.e., either fall into fixed points or evolve to $\pm \infty$,
in the stochastic case they become stable.
By stable we mean the orbits no longer diverge or go into sinks,
instead the velocity function $dx(t)/dt$ may have a complex behavior,
but in such a way to assure that the particle is confined to a
certain limited region of space and do not stop moving.

In conclusion, to understand how the natural laws, so economical
in number and so simple in structure, determine the huge behavioral
richness perceived in the physical world is one of the great
challenges in science
\cite{nakamura-1997,page-2011,rand-1994,rand-1995,stankovski-2017}.
It has been long known that random inputs
\cite{arnold-2002,freidlin-2012},
in otherwise deterministic systems,
can account for part of such multiplicity 
\cite{rubchinsky-1999,bragard-2001,russo-2011,nicolaou-2021,
  wang-2021}.
Although much progress has been achieved, the effects promoting
diversity in noise-assisted dynamical evolution
are very far from a complete description (see, for instance,
\cite{dudkowski-2016,zaslavsky-2002,manrubia-2004,yam-1997}).
We hope that at least for some interesting cases, the present
theoretical results can help to shed some light into this crucial
query.

\section*{Acknowledgments}
We would like to thank G. V. Viswanathan for a critical reading
of earlier versions of the present manuscript and M. W. Beims
for helpful discussions about stabilization processes in
dynamical systems.
MGEL acknowledges financial support from
CAPES (via the CAPES PRINT-UFPR program “Efficiency in uptake,
production and distribution of photovoltaic energy distribution
as well as other renewable sources of renewable energy sources”)
Grant No. 88881.311780/2018-00 and CNPq for the research
Grant No. 304532/2019-3.
YBM acknowledges CAPES for a PhD scholarship. 
CLH is supported in part by the National Science and Technology Council (NSTC) of
the Republic of China under Grant No. NSTC 112-2112-M-032-007.
\appendix

\section{The implicit analytic solution for the dynamical
  system represented by Eqs. (\ref{eq1-det}) and (\ref{eq1a-det})}
\label{appendix1}

Our dynamical system can be solved analytically by considering
the successive spatial intervals $I_n^{(\pm)} = (c_n^{(\pm)},d_n^{(\pm)})$
--- with sign$[G(x \in I_n^{(\pm)})] = \pm 1$ --- to which $x_n(t)$
belongs to at the corresponding time intervals ${\mathcal I}_n$.
The full trajectory $x(t)$ is then given by the proper concatenation
of these piecewise $x_n(t)$ for $t \in {\mathcal I}_n$.

So, here we discuss only the functional form of $x(t)$ in an
arbitrary $I^{(+)}$ or $I^{(-)}$ with $t \in {\mathcal I} =
(t_0,T)$ and for simplicity assuming the staring point $x_0 = x(t_0)$
in the interior of $I^{(\pm)}$.
We observe that in concrete instances, one also should
correctly deal with the behavior of $x(t)$ in crossing
from a spatial interval $I^{(\pm)}$ to $I^{(\mp)}$.
But as seen in Sec. \ref{sec:model-det}, this demands to know
the exact form of $D(x)$.

For the following let us set $G_0 = G(x_0)$ and denote by
$\tilde{G}$ the formal inverse function of $G$.
Hence, for all $x$ it holds that $\tilde{G}(G(x)) = x$.

\subsection{$x(t) \in I^{(+)}$ for the time interval $t \in (t_0,T)$}

In this case $G(x) \geq  0$ (with the equality only at the borders
of $I^{(+)}$) and
\begin{equation}
  \frac{dx(t)}{dt} = c \, G(x)^{\mu-1} \, D(x),
  \label{eq:a1}
\end{equation}
with $c = a - b \, \mu/2 \neq 0$ (of course, $c=0$ leads
to a trivial solution).
Note that $G_0 > 0$.

Then, by the direct integration of Eq. (\ref{eq:a1}) taking into
account Eq. (\ref{eq1a-det}), we find
that ($t \in (t_0,T)$)
\begin{eqnarray}
  x(t) &=& \tilde{G}\big([c \, (2-\mu) \, (t-t_0)
    + G_0^{2 - \mu}]^{1/(2-\mu)}\big),
  \ \ \ \ \  \mbox{for} \ \ \mu \neq 2,
  \nonumber \\
  x(t) &=& \tilde{G}\big(G_0 \, \exp[c \, (t-t_0)]\big),
  \qquad \qquad \qquad \qquad \mbox{for} \ \ \mu = 2.
\end{eqnarray}

\subsection{$x(t) \in I^{(-)}$ for the time interval $t \in (t_0,T)$}

Now $G(x) \leq  0$ (with the equality only at the borders
of $I^{(-)}$) and
\begin{equation}
  \frac{dx(t)}{dt} = d \, [-G(x)]^{\mu-1} \, D(x),
  \label{eq:a2}
\end{equation}
with $d = a + b \, \mu/2 \neq 0$ (again, the $d=0$ case
is trivial).
Observe that $G_0 < 0$.

Finally, by integrating Eq. (\ref{eq:a2}) using
Eq. (\ref{eq1a-det}) we get ($t \in (t_0,T)$)
\begin{eqnarray}
  x(t) &=& \tilde{G}\big(-[d \, (\mu - 2) \, (t-t_0)
    + (-G_0)^{2 - \mu}]^{1/(2-\mu)}\big),
  \ \ \ \, \mbox{for} \ \ \mu \neq 2,
  \nonumber \\
  x(t) &=& \tilde{G}\big(G_0 \, \exp[-d \, (t-t_0)]\big),
  \qquad \qquad \qquad \qquad \qquad \mbox{for} \ \ \mu = 2.
\end{eqnarray}

\end{document}